\bigskip
\documentclass[twocolumn,showpacs,preprintnumbers,amsmath,amssymb]{revtex4}
\usepackage{graphicx}
\usepackage{fancyhdr}
\usepackage{dcolumn}
\usepackage{amssymb}
\usepackage{amsmath}
\usepackage[normalem]{ulem}
\usepackage{bm}
\usepackage{textcomp}
  \usepackage{natbib}
  \usepackage[a4paper,left=1.55cm,top=2cm,right=1.55cm]{geometry}
\usepackage[colorlinks, linkcolor=red, anchorcolor=green,citecolor=blue ]{hyperref}
\usepackage[dvips]{color}

\begin{document}
\preprint{ }
\title{Novel halos in light kaonic nuclei as an indicator of nuclear equation of state at supra-normal densities}
\author{Rong-Yao Yang, Wei-Zhou Jiang, Si-Na Wei, Dong-Rui Zhang}
\affiliation{Department of Physics, Southeast University,
Nanjing  211189, China}

{\begin{abstract} The sensitive correlations between the low-density halo structure and the high-density properties of the nuclear equation of state (EOS) are constructed in light kaonic nuclei with
the relativistic mean-field theory. More specifically, the $1p_{1/2}$ halo spreads out linearly  with increasing the pressure and sound velocity square at supra-normal densities and  decreasing the incompressibility at saturation density. These results suggest that the novel halo in light kaonic nuclei can serve as a sensitive indicator of the nuclear EOS of symmetric matter at supra-normal densities.
\end{abstract}}

\maketitle

\emph{Introduction}.---
The nuclear equation of state (EOS) plays a crucial role in nuclear structures, reaction dynamics and many issues in  astrophysics. As a residual interaction of the quantum chromodynamics, the  nuclear force and the resulting EOS have not been well determined in the medium especially at supra-normal densities due to the complexity of the many-body problem. Even with the nuclear potentials that fit the deuteron properties and nucleon-nucleon scattering data, the nuclear saturation in microscopic approaches does not turn out to be straightforward~\cite{day81,br90,lizh06}.  Moreover, theoretical models that fit the properties of nuclear saturation and finite nuclei yield a large variety of symmetric matter EOS's~\cite{12EK,14JRS,14MD} and  diverse density dependence of symmetry energy~\cite{05AWS,06FUC,08BAL} both of which differ largely at supra-normal densities. In terrestrial laboratories~\cite{85JA,02PD,06CF,08BAL,09MBT,12CH}, the energetic heavy-ion reactions are currently the unique way to determine the high-density EOS, while the celestial observation of neutron stars (NS's) may provide another hopeful way to constrain the high-density EOS~\cite{98AA,04JML,07JML,10AWS,10PBD,12AWS,13AWS,13KH}. However, the extracted EOS suffers from large uncertainties  that are close to a large relative error of 50\% at high densities~\cite{02PD,08BAL,10AWS}.  Moreover, significant uncertainties of the celestial constraints may also arise from the possible dark matter contamination in NS's~\cite{11PC,14XQF} and potential deviation from the standard Einstein theory in strong-field limits, e.g., see Ref.~\cite{15XTH}. Although  the elementary forces and the structure of matter are two basic subjects in physics, such a large systematic error prohibits indeed from extracting  structural properties such as phase transitions and matter constituents. Thus, it is of prime importance and broad interest  to pursuit accurate extraction of the nuclear EOS at supra-normal densities.

It is well known that  the structural properties of finite nuclear system may accurately  constrain the nuclear forces and the resulting nuclear EOS near or beneath saturation density. Recently, the properties of exotic nuclei have played a special role in constraining the nuclear forces. For instance, the exotic states, such as the halo~\cite{85IT,93MVZ,13HWH} and Hoyle states~\cite{06WVO,11EEP}, and the shell evolution anomaly~\cite{10TO,10TOS} reveal the importance of the three-body force, while the novel magic numbers far off the $\beta$-stability are mainly subject to the role of the tensor force~\cite{01TO,05TO,06OS}.  It is, however, unfortunate  that  the high-density EOS can not be determined by the structural properties of finite nuclei, since the extrapolation of the EOS's diversifies greatly at high densities.  Thus,  novel systems that feature a much compacter core should be created to  constrain the high-density EOS directly. While it is impossible to acquire a clear rise of the core density by adding more nucleons in finite nuclei, the inclusion of the new degree of freedom such as the strangeness becomes a uniquely possible  way to realize a much denser core in finite nuclei because of the additional attraction. Typical examples are the metastable exotic multihypernuclear objects~\cite{93JS,94JS,06WZJ} and  kaonic nuclei~\cite{99TK,02YA,02TY}. In particular, due to the strongly attractive interaction between nucleons and $K^-$ meson, the deep $K^-$-nuclear bound states may form,  resulting in a high-density core in light kaonic nuclei~\cite{02YA,02TY,06JM,06XHZ,07DG,14RYY}. Up to now, continuous experimental efforts have been made progressively to search for kaonic nuclei~\cite{05MA,06VKM,07GB,10TY,13SA,14AOT,15YI,15AF,15GA}.
In this letter, we reveal noticeably  for the first time that the strangeness in deeply bound kaonic nuclei may provide a novel mechanism for the formation of the exotic structure, the diffusive nuclear halo.
This brand-new mechanism   enables the halo formation in nuclei of $\beta$-stability, while the normal halos are usually restrained to the neighborhood of the drip lines~\cite{98MJ,05CAB,06MJ}.
We will find that the property of the low-density halo correlates sensitively with the incompressibility at saturation density and the pressure and sound velocity at supra-normal densities. These relationships arising from the different density regimes in the same system will provide accurate constraints on the high-density EOS.

\emph{Formalism}.---
\label{RMF}
In this work, the light kaonic nuclei are studied in the relativistic mean-field (RMF) models~\cite{Wal74,Boguta77,Ser86,Ring96}. The interacting Lagrangian is written as~\cite{01HO,05BGT}
\begin{eqnarray} \label{EL0}
\mathcal{L}_{int}&=& \bar{\psi}_N[ g_{\sigma N}\sigma - g_{\omega N}\gamma_{\mu}\omega^{\mu}
 - g_{\rho N}\gamma_{\mu}\tau_3 b_0^{\mu}-  \nonumber\\
&& e\frac{1+\tau_3}{2}\gamma_{\mu}A^{\mu}]\psi_N
 - \frac{1}{3}g_2\sigma^3
- \frac{1}{4}g_3\sigma^4  \nonumber\\
&& +\frac{1}{4}c_3(\omega_\mu\omega^\mu)^2.
\end{eqnarray}
The Lagrangian density includes the interactions between the nucleon field and three meson fields: an isoscalar-scalar  $\sigma$, an isovector-vector $b_0^\mu$ and a vector $\omega_\mu$, the Coulomb interaction $A_\mu$, and the nonlinear self-interactions of meson fields. The meson self-interactions are included to adjust the incompressibility and the stiffness of the EOS in the high-density region.
In the RMF approximation, the nuclear EOS including the energy density and the pressure can be derived from the Lagrangian density~\cite{Ser86}. The nucleon potential in the Dirac equation reads
\begin{equation} \label{Epot}
U_N = -g_{\sigma
N}\sigma_0+ g_{\omega N}\omega_0 + g_{\rho N}\tau_3
b_0+e\frac{1+\tau_3}{2}A_0.
\end{equation}

The Lagrangian of the $K^-$N interaction is given by~\cite{07DG}
\begin{equation} \label{ELKN}
\mathcal{L}_{KN} = (\mathcal{D}_{\mu}K)^{\dag}(\mathcal{D}^{\mu}K)
 - (m^2_{K} - g_{\sigma K} m_{K} \sigma )K^{\dag}K,
\end{equation}
where the covariant derivative is expressed as
\begin{equation} \label{Eder}
 \mathcal{D}_{\mu} \equiv \partial _{\mu} + i g_{\omega K} \omega_{\mu} + i g_{\rho K}
 b_{0\mu} + ie\frac{1+\tau_3}{2}A_{\mu},
\end{equation}
with the $g_{iK},$ $(i=\sigma, \omega, \rho)$ being the corresponding
meson-$K^-$ coupling constants and $m_K$ is the mass of $K^-$ meson. Here, the $K^-$N interaction is mediated by $\sigma-K^-$, $\omega-K^-$, $\rho-K^-$ and photon-$K^-$ couplings.  The resulting Dirac equation for nucleons and Klein-Gordon equation for $K^-$ meson, obtained from the Lagrangian in the RMF approximation, are   solved in a self-consistently iterative way~\cite{14RYY}.

\emph{Results and discussions}.---
\label{result}
An interesting property of meson mediated interactions for $K^-$ is that they are coherently attractive, giving rise to the strongly attractive $K^-$N interaction. This is supported by fitting the kaonic atom data~\cite{81batty,94EF,97CJB,99EF,01gal,07EF} or by analyzing the low-energy $K^-$N scattering data based on chirally motivated models~\cite{96waas,97waas,97JSB,00AR,01AC,11AC}. From the former, the $K^-$ optical  potential may reach as deep as -180MeV  at saturation density~\cite{94EF,97CJB,99EF}, and from the latter the strong attraction can also be remarkable with a deep depth of -120MeV~\cite{96waas,97waas}. It should be noted that the applicability of  chirally motivated models should be restrained by the chiral dynamics to low densities. Whereas both methods need some specific extrapolations to  saturation density, diversification appears with the much shallower depth around -50MeV in both methods~\cite{81batty,01gal,00AR,01AC}. Fortunately, the heavy-ion collisions can provide a direct way  to extract the $K^-$ potential at appropriately produced densities. In the past, various heavy ion collisions got the almost consistent $K^-$ potential depth around -100 MeV at saturation density~\cite{97GQL,99WC}.  Recently, the direct approval of  the  strongly attractive  $K^-$N interaction was provided one more time  by the deep $K^-$ optical potential depth, -100 MeV  at saturation density~\cite{14ZQF}, extracted from the collision data from the KaoS Collaboration~\cite{99FL,03AF,06WS}. In the study to reveal the nuclear EOS effects  on the nuclear halo, we thus use the $K^-$ optical potential depth of -100 MeV  at saturation density. Eventually, the effect of various $K^-$ optical potential depthes on the halo will be concerned.
The coupling constants $g_{\rho K}$ and $g_{\omega K}$ are determined by the SU(3) relation: $2g_{\omega K}=2g_ { \rho K}=g_ { \rho \pi}=6.04$, while the unique free parameter, $g_{\sigma K}$, is adjusted to fit the depth of $K^-$ optical potential~\cite{14RYY}.

\begin{figure}[thb]
\centering
\includegraphics[height=7.0cm,width=6cm]{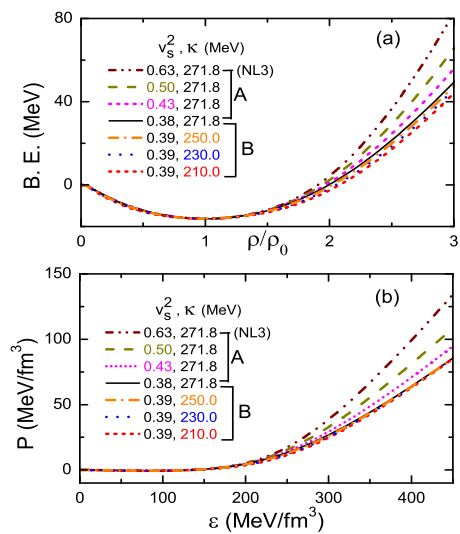}
\caption{(Color online) The binding energy per nucleon (B. E.) as a function of nuclear density (upper panel) and the relationship between the pressure and energy density (lower panel) for various EOS's of symmetric matter. The sound velocity square $v_s^2$ at 2.5$\rho_0$ and incompressibility $\kappa$ at $\rho_0 (= 0.148 fm^{-3})$ are denoted for each curve. } \label{Feos}
\end{figure}

We simulate the uncertainty of the nuclear EOS by adjusting  the strengthes of the self-interacting terms in Eq.(\ref{EL0}) in a traditional way~\cite{Boguta77,05BGT}, while the other parameters in $\mathcal{L}_{int}$ are just moderately modified (less than 5\%). As a result, the various stiffness of the EOS is produced at supra-normal densities or different incompressibility produced at saturation density ($\rho_0$). Here, we take the famous NL3 parameter set as a starting point~\cite{97GAL}, like the parametrization work in Ref.~\cite{05BGT}. Fig.~\ref{Feos} shows different nuclear EOS's given in two schemes. In Scheme A, we soften the nuclear EOS in the high-density region but keep the saturation property unchanged. Such EOS softening, denoted in Fig.~\ref{Feos} by the descending sound velocity square $v_s^2$ with  $v_s^2=\partial P/\partial\epsilon$  being the partial derivative of the pressure with respect to the energy density, can be realized by increasing the $\omega$ meson self-interacting coupling $c_3$ and correspondingly adjusting the parameters of the $\sigma$ meson self-interacting terms.  The Scheme B is given to determine  a particular incompressibility $\kappa$ at $\rho_0$ by modifying mainly the parameters of the  $\sigma$ meson self-interacting terms while with the fixed $c_3$=60. This yields a series of EOS's with a similar high-density behavior but with the different incompressibility. These two schemes enable us to investigate separately the effects of the high-density EOS stiffness and of the saturation property on the formation of the nuclear halo. The relevant parameter sets concerning the two schemes are given in Table~\ref{Tpara}.

\begin{table}[htp]
\caption{ The parameters for various nuclear EOS's and the $g_{\sigma K}$ used in Scheme A (upper rows) and B (lower rows). The unlisted parameters are the same as those of the parameter set NL3. $g_2$ and $m_\sigma$ are in unit of $fm^{-1}$ and MeV, respectively. Also given are the sound velocity square $v_s^2$ at 2.5$\rho_0$ and the incompressibility  $\kappa$ (in unit of MeV).
\label{Tpara}}
\begin{center}
\begin{tabular}{c | c | c | c | c | c | c || c | c }
\hline\hline

 $c_3$ &  $g_2$ & $g_3$ & $g_{\sigma N}$ & $g_{\omega N}$  & $g_{\sigma K}$ & $m_\sigma$ & $v_s^2$ & $\kappa$\\
\hline
 0  & 10.431& -28.885 & 10.217 & 12.868 & 1.455 & 508.2 & 0.63 & 271.8\\
  20 & 9.833 & -23.290 & 10.274 & 12.985 & 1.496 & 508.2 & 0.50 & 271.8 \\
 40 & 9.205 & -17.781 & 10.323 & 13.097 & 1.535  & 508.2 & 0.43 & 271.8 \\
 60 & 8.604 & -12.444 & 10.370 & 13.205 & 1.572  & 508.2 & 0.38 & 271.8 \\
 90 & 7.750 & -4.731  & 10.436 & 13.359 & 1.624  & 508.2 & 0.34 & 271.8\\
\hline
 60 & 11.369 & -22.987 & 10.164 & 13.205  & 1.541 & 490.0 & 0.39 & 210.0\\
 60 & 10.479 & -19.702 & 10.214 & 13.205  & 1.548 & 495.0 & 0.39 & 230.0 \\
 60 & 9.656 & -16.472 & 10.303 & 13.205  & 1.562 & 502.0 & 0.39 & 250.0 \\
 60 & 7.695& -8.875 & 10.437 & 13.205  & 1.582 & 514.0 & 0.38 & 290.0 \\
\hline
\hline
\end{tabular}
\end{center}
\end{table}

\begin{figure}[thb]
\centering
\includegraphics[height=6cm,width=7.6cm]{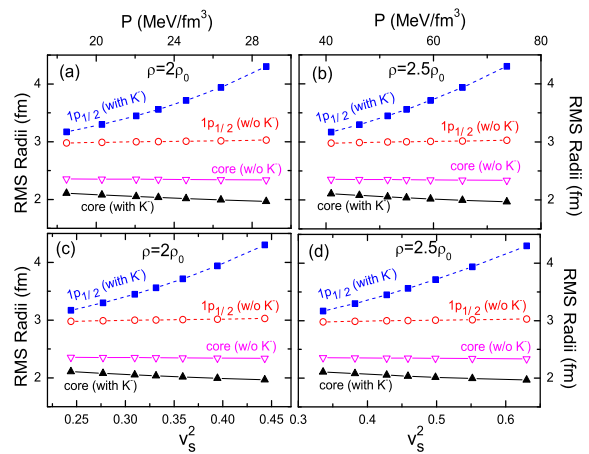}
\caption{(Color online) The RMS radii of the core and the outmost layer neutron in $^{13}$C and $^{13}_{K^-}$C as a function of the pressure (up panels) and  sound velocity square (lower panels) at $\rho$=2$\rho_0$ (left  panels) and 2.5$\rho_0$ (right panels) for various EOS's in Scheme A. } \label{Frms1}
\end{figure}

\begin{figure}[thb]
\centering
\includegraphics[height=4.5cm,width=6cm]{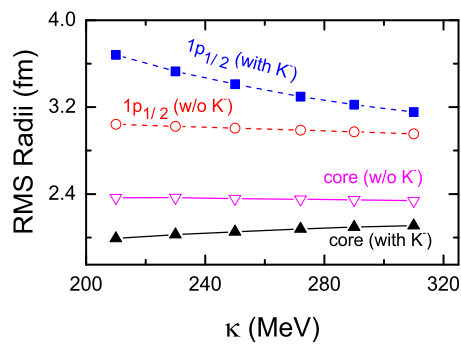}
\caption{(Color online)  The RMS radii of the core and the outmost layer neutron in $^{13}$C and $^{13}_{K^-}$C as a  function of the incompressibility  $\kappa$ with various $\kappa$ obtained in Scheme B. } \label{Frms2}
\end{figure}

With the parameter sets that give rise to the nuclear EOS's of symmetric matter shown in Fig.~\ref{Feos}, we investigate the halo phenomenon in light kaonic nuclei that feature the outmost layer nucleons in the $1p_{1/2}$ orbital. Here, we take $^{13}$C as the seed nucleus for the $K^-$ implantation as an example. Fig.~\ref{Frms1} displays  root-mean-square (RMS) radii of the  core  $^{12}$C and the $1p_{1/2}$ neutron in $^{13}$C and $^{13}_{K^-}$C as a function of the pressure and the sound velocity square at $\rho$=2$\rho_0$  and $2.5\rho_0$ which are the densities reachable within the kaonic nuclei, see Table~\ref{T1}.  As shown in Fig.~\ref{Frms1}, the radius of the ${1p_{1/2}}$ neutron in $^{13}_{K^-}$C increases very significantly with the stiffening of the EOS at supra-normal densities, while all other radii  including the radius of the $1p_{1/2}$ neutron in normal $^{13}$C are insensitive to the variation of the high-density EOS.
With the stiffening of the high-density EOS, a diffusive neutron halo thus  forms in $^{13}_{K^-}$C, while there is anyway no  halo phenomenon in normal $^{13}$C, in accord with experiments. Meanwhile, as seen from  Table~\ref{T1}, the core radius of $^{13}_{K^-}$C is clearly smaller than that of normal $^{13}$C due to the shrinkage induced by the strong $K^-N$ attraction. Interestingly, the correlation between the radius of the $1p_{1/2}$ halo neutron and the pressure and sound velocity square is nearly linear at supra-normal densities, especially at a higher density. Analogously, the correlation between the radii of the core and outmost layer neutron and the  incompressibility at saturation density can be established, as shown in Fig.~\ref{Frms2}. We observe that  the radius of the $1p_{1/2}$ halo neutron decreases almost linearly with the rise of  the incompressibility. These results provide us a striking perspective to constrain the high-density EOS through the correlation with the low-density halo in the same nutshell, which lowers greatly the large uncertainty of the conventional extrapolation method.

\begin{table*}[htp]
\caption{ Single-neutron binding energies, the core and $1p_{1/2}$ neutron radii $R_{c}$ and $R_{h}$, and the maximum nuclear density in $^{13}_{K^-}$C with various EOS's. The columns denoted by the $v_s^2$ (at 2.5$\rho_0$) and  $\kappa$  values correspond to Scheme A and B, respectively. The column without $K^-$ is obtained  with the NL3.  The  binding energies and radii are in unit of MeV and fm, respectively.
\label{T1}}
\begin{center}
\begin{tabular}{ c || c | c | c | c | c | c | c | c  }
\hline\hline

& w/o $K^-$ & $v_s^2$=0.63 & $v_s^2$=0.50 &  $v_s^2$=0.43 &  $v_s^2$=0.38 & $\kappa$=250.0 & $\kappa$=230.0 & $\kappa$=210.0   \\
\hline
$R_{c}$ & 2.34 & 1.97 & 2.02 & 2.05 & 2.08 & 2.05 & 2.03 & 1.99   \\
$R_{h}$ & 3.16 & 4.30 & 3.72 & 3.45 & 3.30 & 3.41 & 3.53 & 3.68  \\
$\rho_{Max}/\rho_0$ & 1.53 & 2.53 &	2.46 & 2.41 & 2.35 & 2.45 &	2.54 & 2.62   \\
\hline
$1s_{1/2}$ & 43.90 & 87.10 & 74.99 & 68.12 & 63.99 & 68.25 & 72.91 & 79.25   \\
$1p_{3/2}$ & 17.90 & 25.23 & 23.83 & 22.93 & 22.36 & 22.96 & 23.59 & 24.54   \\
$1p_{1/2}$ & 8.45  & 1.41  & 2.90  & 4.12  & 5.05  & 4.42  & 3.87 & 3.21   \\

\hline
\hline
     \end{tabular}
     \end{center}
  \end{table*}

\begin{figure}[thb]
\includegraphics[height=6.0cm,width=6cm]{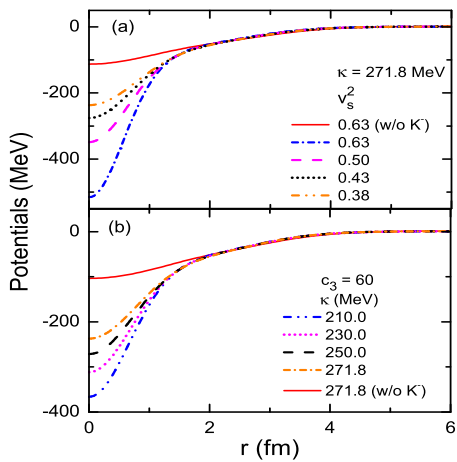}
\caption{(Color online) The nucleon potential, given by Eq.(\ref{Epot}), as a function of radius with various nuclear EOS's in $^{13}$C and $^{13}_{K^-}$C: (a) with the given $\kappa$=271.8 MeV (Scheme A), and (b) with the given $c_3=60$ (Scheme B). } \label{Fpot}
\end{figure}

To reveal the physics why the halo structure in kaonic nuclei is sensitive to the EOS, we investigate the nucleon potential, as shown in Fig.~\ref{Fpot}. It is  seen  that the implantation of  the $K^-$ meson deepens the nuclear potential greatly. The deepening role is  clearly strengthened either due to the  enhancement of the $K^-$-nucleon attraction by  stiffening  the  EOS  at supra-normal densities (with larger $v_s^2$ and resulting  stiffer vector field $\omega_0$, panel a) or by ensuring easier compression  (with smaller $\kappa$ at $\rho_0$, panel b).  As a result, the clear separation in the nucleon potential for various EOS's yields rather distinct ground-state properties of kaonic nuclei.
Indeed, we can see  from Table~\ref{T1} that the separation between neighboring neutron energy levels in $^{13}_{K^-}$C becomes progressively large with increasing $v_s^2$ or reducing $\kappa$. This phenomenon can roughly be understood in a simple quantum model  with the harmonic potential  where stiffer potentials give larger level separation, though the mean-field potential in $^{13}_{K^-}$C is a little different from harmonic potential. Since the deep potential well is  also in resemblance to the infinite deep square potential well, we can have the similar understanding of the large level separation caused by the stiffening of the high-density EOS or the reduction of the incompressibility.
Since the single-neutron  energy of $1p_{3/2}$ state in $^{13}_{K^-}$C is only slightly affected by various EOS's, the increasing spin-orbit splitting of the $1p$ orbitals is dominated by the shift of the  $1p_{1/2}$ state.  By deepening the nucleon potential, the  $1p_{1/2}$  neutron becomes closer to  the continuum, resulting in the birth of the halo structure with a diffusive extension of the last neutron  in $^{13}_{K^-}$C. Given the more and more constrained incompressibility, see~\cite{14JRS,99Young} and references therein,   the stiffness of the high-density EOS should dominate  the halo structure.   Meanwhile, the core radius of $^{13}_{K^-}$C becomes shrunk because of the deep binding of interior states in $^{13}_{K^-}$C. The shrinkage induced by the strong $K^-N$ attraction gives rise to a large maximum density that can reach up to 2.62$\rho_0$.
Note that the nucleon potentials with various parametrizations in normal $^{13}$C are quite similar with each other, consistent with the proximity in their RMS radii and single-particle properties. Besides, the parametrizations with various symmetry energies just have negligible  difference in the halo in nuclei of $\beta$-stability since the isovector potential is small,  compared with the deep nucleon potential well.

It is worthy to point out that the sensitive dependence of the halo structure on the nuclear EOS is also discovered in other light kaonic nuclei that possess the  nucleon occupation of the outmost $1p_{1/2}$ orbital, e.g., $^{14}_{K^-}$C, $^{14}_{K^-}$O, and etc. Moreover, we find that the conclusions are qualitatively the same with the $K^-$ optical potential depth ranging from  -80  to -120 MeV which is coincident with the range extracted from the proton-nucleus and nucleus-nucleus collisions~\cite{97GQL,99WC,06WS,99FL,03AF,14ZQF}. Within this  range, the imaginary part of $K^-$ optical potential just has minor effects on properties of kaonic nuclei and is  neglected here~\cite{06JM,14RYY,15xu}. In addition, we have examined the model dependence of the correlation relationships.  The simulation based on the RMF parameter set TM2~\cite{Sug94}, rather different from the NL3, indicates that the sensitive correlation between  the nuclear EOS and the halo preserves and is rather model independent. The present results suggest that these light kaonic nuclei can be used as the favorable candidates to constrain  the nuclear EOS at supra-normal densities either with various theoretical approaches or by performing the experiments. As an experimental suggestion, the photo-nucleus or pion-nucleus reaction~\cite{Kaiser97,14AOT}, e.g., $\gamma$ + $^{13}$C $\rightarrow$ $K^+$  +  $^{13}_{K^-}$C or $\pi^-$ + $^{13}N$ $\rightarrow$ $K^{*+}$ + $^{13}_{K^-}$C, may be used to produce kaonic nuclei, with an anticipation of the halo radius measurement via the correlation of the outgoing kaons with the halo neutron due to the strong interaction.

\emph{Summary}.---
\label{summary}
In this work, we have investigated in the RMF theory the novel halo formation due to the strong $K^-N$ attraction  in light kaonic nuclei that feature the outmost nucleons in the $1p_{1/2}$ orbital.  It is found that the low-density halo radius correlates very sensitively with the nuclear EOS of symmetric matter at saturation density and in the region of supra-normal densities which forms in the core of light kaonic nuclei. Facing up large uncertainties of the nuclear EOS at supra-normal densities either due to extrapolations with any nuclear model or from the extractions through heavy-ion reactions or celestial observations of neutron stars, the present method with the structural exploration  has the appealing merit to evade from such uncertainties. In particular, the determination of the high-density EOS can be implemented through the nearly linear correlation between the property of low-density diffusive halos and the pressure and sound velocity square in the region with a density up to $2.5\rho_0$.  We hope that the present study may urge more theoretical explorations and especially the timely experiments for light kaonic nuclei.

We are indebted to Profs. Qiang Zhao, Zhao-Qing Feng,  Tomofumi Nagae for helpful discussions. The work was
supported in part by the National Natural Science Foundation of China
under Grant No. 11275048,  the China Jiangsu Provincial Natural
Science Foundation under Grant No.BK20131286.


\begin{thebibliography}{99}

\bibitem{day81}B. D. Day, Phys. Rev. Lett., \textbf{47}, 226 (1981).
\bibitem{br90}R. Brockmann and  R. Machleidt,  Phys. Rev. C \textbf{42}, 1965 (1990).
\bibitem{lizh06}Z. H. Li, U. Lombardo, H. J. Schulze, W. Zuo, L. W. Chen, and H. R. Ma,  Phys. Rev. C \textbf{74}, 047304 (2006).

\bibitem{12EK} E. Khan, J. Margueron, and I. Vidana, Phys. Rev. Lett. \textbf{109}, 092501  (2012).
\bibitem{14JRS} J. R. Stone, N. J. Stone, and S. A. Moszkowski, Phys. Rev. C \textbf{89}, 044316 (2014).
\bibitem{14MD} M. Dutra, O. Lourenco, S. S. Avancini et al., Phys. Rev. C \textbf{90}, 055203 (2014).

\bibitem{05AWS} A. W. Steiner, M. Prakash, J. M. Lattimer, and P. Ellis, Phys. Rep. \textbf{411}, 325 (2005).
\bibitem{06FUC} C. Fuchs and H. H. Wolter,  Eur. Phys. J. A \textbf{30}, 5 (2006).
\bibitem{08BAL} B. A. Li, L. W. Chen, and C. M. Ko, Phys. Rep. \textbf{464}, 113 (2008).

\bibitem{85JA} J. Aichelin and C. M. Ko, Phys. Rev. Lett. \textbf{55}, 2661 (1985).
\bibitem{02PD} P. Danielewicz, R. Lacey, and W. G. Lynch, Science \textbf{298}, 1592 (2002).
\bibitem{06CF}C. Fuchs, Prog. Part. Nucl. Phys. \textbf{56}, 1 (2006).
\bibitem{09MBT} M. B. Tsang, Y. X. Zhang, P. Danielewicz et al., Phys. Rev. Lett. \textbf{102}, 122701 (2009).
\bibitem{12CH} C. Hartnack, H. Oeschler, Y. Leifels et al., Phys. Rep. \textbf{510}, 119 (2012).

\bibitem{98AA} A. Akmal, V. R. Pandharipande, and D. G. Ravenhall, Phys. Rev. C \textbf{58}, 1804 (1998).
\bibitem{04JML} J. M. Lattimer and M. Prakash, Science \textbf{304}, 536 (2004).
\bibitem{07JML} J. M. Lattimer and M. Prakash, Phys. Rep. \textbf{442}, 109 (2007).
\bibitem{10PBD} P. B. Demorest, T. Pennucci, S. M. Ransom et al., Nature \textbf{467}, 1081 (2010).
\bibitem{10AWS} A. W. Steiner, J. M. Lattimer, and E. F. Brown, Astrophys. J. \textbf{722}, 33 (2010).
\bibitem{12AWS} A. W. Steiner and S. Gandolfi, Phys. Rev. Lett. \textbf{108}, 081102 (2012)
\bibitem{13AWS} A. W. Steiner, J. M. Lattimer, and E. F. Brown, Astrophys. J. Lett. \textbf{765}, L5 (2013).
\bibitem{13KH} K. Hebeler, J. M. Lattimer, C. J. Pethick, and A. Schwenk, Astrophys. J. \textbf{773}, 11 (2013).

\bibitem{11PC} P. Ciarcellut and F. Sandin, Phys. Lett. B \textbf{695}, 19 (2011).
\bibitem{14XQF} Q. F. Xiang, W. Z. Jiang, D. R. Zhang, and R. Y. Yang, Phys. Rev. C \textbf{89}, 025803 (2014).

\bibitem{15XTH}X. T. He, F. J. Fattoyev, B. A. Li, W. G. Newton, Phys. Rev. C \textbf{91}, 015810 (2015).
\bibitem{85IT} I. Tanihata, H. Hamagaki, O. Hashimoto et al., Phys. Rev. Lett. \textbf{55}, 2676 (1985).
\bibitem{93MVZ} M. V. Zhukov, B. V. Danilin, D. V. Fedorov et al., Phys. Rep. \textbf{231}, 151 (1993).
\bibitem{13HWH}H. W. Hammer, A. Nogga, A. Schwenk, Rev. Mod. Phys. \textbf{85}, 197 (2013).
\bibitem{06WVO}W. von Oertzen, M. Freer, and Y. Kanada-En'yo, Phys. Rep. \textbf{432}, 43 (2006).
\bibitem{11EEP} E. Epelbaum,  H. Krebs, D. Lee, and Ulf-G. Meissner, Phys. Rev. Lett. \textbf{106}, 192501 (2011).
\bibitem{10TO}T. Otsuka, T. Suzuki, M. Honma, et. al., Phys. Rev. Lett. \textbf{104}, 012501 (2010).
\bibitem{10TOS}T. Otsuka, T. Suzuki, J. D. Holt, A. Schwenk, and Y. Akaishi, Phys. Rev. Lett. \textbf{105}, 032501 (2010).
\bibitem{01TO}T. Otsuka, R. Fujimoto, Y. Utsuno, B. A. Brown, M. Honma, T. Mizusaki Phys. Rev. Lett. \textbf{87}, 082502 (2001).
\bibitem{05TO}T. Otsuka, T. Suzuki, R. Fujimoto,  H. Grawe, and Y. Akaishi, Phys. Rev. Lett., \textbf{95}, 232502 (2005).
\bibitem{06OS}O. Sorlin, M. G. Porquet, Prog. Part. Nucl. Phys., \textbf{61}, 602, (2008).

\bibitem{93JS}J. Schaffner, C. B. Dover, A. Gal, C. Greiner, H. St\"oker, Phys. Rev. Lett. {\bf 71}, 1328 (1993)
\bibitem{94JS} J. Schaffner, C. B. Dover, A. Gal, C. Greiner, D. J. Millener, H. St\"oker, Ann. Phys. {\bf 235}, 35 (1994).
\bibitem{06WZJ} W. Z. Jiang, Phys. Lett. B \textbf{642}, 28 (2006).
\bibitem{99TK} T. Kishimoto, Phys. Rev. Lett. \textbf{83}, 4701 (1999).
\bibitem{02TY} T. Yamazaki, and Y. Akaishi, Phys. Lett. B \textbf{353}, 70 (2002).
\bibitem{02YA}Y. Akaishi, and T. Yamazaki, Phys. Rev. C \textbf{65}, 044005 (2002).

\bibitem{06JM} J. Mare\v{s}, E. Friedman, and A. Gal, Nucl. Phys. A \textbf{770}, 84 (2006).
\bibitem{06XHZ} X. H. Zhong, G. X. Peng, L. Li, and P. Z. Ning, Phys. Rev. C \textbf{74}, 034321 (2006).
\bibitem{07DG}D. Gazda, E. Friedman, A. Gal, and J. Mare\v{s}, Phys. Rev. C \textbf{76}, 055204 (2007).
\bibitem{14RYY} R. Y. Yang, W. Z. Jiang, D. R. Zhang, S. N. Wei, Eur. Phys. J. A \textbf{50}, 1 (2014).

\bibitem{05MA}M. Agnello, \emph{et al}. (FINUDA Collaboration), Phys. Rev. Lett. \textbf{94}, 212303 (2005).
\bibitem{06VKM} V. K. Magas, E. Oset, A. Ramos, and H. Toki, Phys. Rev. C \textbf{74}, 025206 (2006).
\bibitem{07GB} G. Bendiscioli, T. Bressani, A. Fontana et al., Nucl. Phys. A \textbf{789}, 222 (2007).
\bibitem{10TY} T. Yamazaki, \emph{et al}. (DISTO Collaboration), Phys. Rev. Lett. \textbf{104}, 132502 (2010).
\bibitem{13SA} S. Ajimura et al., Nucl. Phys. A \textbf{914}, 315 (2013).
\bibitem{14AOT} A. O. Tokiyasu, \emph{et al}. (LEPS Collaboration), Phys. Lett. B \textbf{728}, 616 (2014).
\bibitem{15YI} Y. Ichikawa, \emph{et al}., Prog. Theor. Exp. Phys. \textbf{2015}, 021D01 (2015).
\bibitem{15AF} A. Filippi and S. Piano, Hyperfine Interact. \textbf{233}, 151 (2015).
\bibitem{15GA} G. Agakishiev, \emph{et al}. (HADES Collaboration), Phys. Lett. B \textbf{742}, 242 (2015).

\bibitem{98MJ}J. Meng and P. Ring, Phys. Rev. Lett. \textbf{80}, 460  (1998).
\bibitem{05CAB}C. A. Bertulani, Phys. Rev. Lett. \textbf{94}, 072701 (2005).
\bibitem{06MJ}J. Meng, H. Toki, S. G. Zhou, S. Q. Zhang, W. H. Long, and
    L. S. Geng, Prog. Part. Nucl. Phys. \textbf{57}, 470 (2006).

\bibitem{Wal74}J. D. Walecka, Ann. Phys.(NY) \textbf{83}, 491 (1974).
\bibitem{Boguta77}J. Boguta and A. R. Bodmer, Nucl. Phys. A \textbf{292}, 413 (1977).
\bibitem{Ser86}B. D. Serot and J. D. Walecka, Adv. Nucl. Phys. \textbf{16}, 1 (1986).
\bibitem{Ring96}P. Ring, Prog. Part. Nucl. Phys. \textbf{37}, 193 (1996).

\bibitem{01HO} C. J. Horowitz and J. Piekarewicz, Phys. Rev. Lett. {\bf 86}, 5647 (2001).
\bibitem{05BGT}B. G. Todd-Rutel and J. Piekarewicz, Phys. Rev. Lett. \textbf{95}, 122501
(2005).

\bibitem{81batty} C. J. Batty, Nucl. Phys. A 372, 418 (1981).
\bibitem{94EF}E. Friedman, A. Gal, and C. J. Batty, Nucl. Phys. A \textbf{579}, 518 (1994).
\bibitem{97CJB} C. J. Batty, E. Friedman, and A. Gal, Phys. Rep. \textbf{287}, 385 (1997).
\bibitem{99EF}E. Friedman, A. Gal, J. Mare\v{s}, and A. Ciepl\'{y}, Phys. Rev. C \textbf{60}, 024314 (1999).
\bibitem{01gal} A. Gal, Nucl. Phys. A \textbf{691}, 268C (2001).
\bibitem{07EF}E. Friedman, and A. Gal, Phys. Rep. \textbf{452}, 89 (2007).

\bibitem{97JSB} J. Schaffner-Bielich, I. N. Mishustin, and J. Bondorf, Nucl. Phys. A \textbf{625}, 325 (1997).
\bibitem{00AR} A. Ramos, and E. Oset, Nucl. Phys. A \textbf{671}, 481 (2000).
\bibitem{01AC} A. Ciepl\'y, E. Friedman, A. Gal, and J. Mare\v{s}, Nucl. Phys. A \textbf{696}, 173 (2001).
\bibitem{11AC} A. Ciepl\'y, E. Friedman, A. Gal, D. Gazda, J. Mare\v{s}, Phys. Lett. B \textbf{702}, 402 (2011).

\bibitem{96waas} T. Waas, N. Kaiser, and W. Weise, Phys. Lett. B \textbf{365}, 12 (1996).
\bibitem{97waas} T. Waas, M. Rho, and W. Weise, Nucl. Phys. A \textbf{617}, 449 (1997).

\bibitem{97GQL} G. Q. Li, C. H. Lee, and G. E. Brown, Phys. Rev. Lett. \textbf{79}, 5214 (1997).
\bibitem{99WC} W. Cassing and E. L. Bratkovskaya, Phys. Rep. \textbf{308}, 65 (1999).

\bibitem{14ZQF} Z. Q. Feng, W. J. Xie, and G. M. Jin, Phys. Rev. C \textbf{90}, 064604 (2014).
\bibitem{99FL}F. Laue, \emph{et al}., Phys. Rev. Lett. \textbf{82}, 1640 (1999).
\bibitem{03AF}A. F\"orster, \emph{et al}. (KaoS Collaboration), Phys. Rev. Lett. \textbf{91}, 152301 (2003).
\bibitem{06WS} W. Scheinast, \emph{et al}. Phys. Rev. Lett. \textbf{96}, 072301 (2006).

\bibitem{97GAL}G. A. Lalazissis, J. Konig, and P. Ring, Phys. Rev. C \textbf{55}, 540 (1997).

\bibitem{99Young}D. H. Youngblood, H. L. Clark, and Y. W. Lui, Phys. Rev. Lett. \textbf{82}, 691 (1999).
\bibitem{15xu} R. L. Xu, C. Wu, W. L. Qian, and Z. Z. Ren, Eur. Phys. J. A \textbf{51}, 20 (2015).
\bibitem{Sug94}Y. Sugahara and H. Toki, Nucl. Phys. A \textbf{579}, 557 (1994).
\bibitem{Kaiser97}N. Kaiser, T. Waas, W. Weise, Nucl. Phys. A \textbf{612}, 297 (1997).

\end{thebibliography}
\end{document}